\documentclass{an}
\usepackage{graphicx}
\usepackage{times}
\usepackage{natbib}
\begin{document}

\Pagespan{1}{}
\Yearpublication{2007}%
\Yearsubmission{2007}%
\Month{}%
\Volume{}%
\Issue{}%

\title{Polarimetric imaging of interacting pairs
\thanks{Based on observations obtained at ESO-La Silla, Chile 
(Observing Runs 074.B-0276(A) and 076.B-0046(A))}}

\author{Roberto Rampazzo
\thanks{Corresponding author:
  \email{roberto.rampazzo@oapd.inaf.it}\newline}
\and  Carlotta Bonoli
\and Enrico Giro
}
\titlerunning{Polarimetric imaging of interacting pairs}
\authorrunning{R. Rampazzo, C. Bonoli \& E. Giro}
\institute{
INAF-Osservatorio Astronomico di Padova, 
Vicolo dell'Osservatorio 5, 
35122 Padova, Italy}

\received{December, 2007}
\accepted{}
\publonline{later}


\abstract{
We present optical polarization maps of a sample of four interacting pairs 
at different {\it phases} of encounter, from nearly 
unperturbed galaxies to on-going mergers. \\
Only the pair RR~24 shows a linear polarization pattern  which
extends in both galaxies for several kiloparsecs. The more perturbed member,
RR~24b, is lineraly polarized up to the level of $\approx$3\%. 
No polarization is measured in the strongly perturbed late-type pair 
members of RR~23 and RR~99. Also, in the central part of the double
nuclei shell galaxy ESO 2400100 there is no significant polarization.\\
We use the ionized gas velocity field of RR~24 to interpret its
linear polarization structure. In RR~24a the quite regular gas kinematics reflect
the unperturbed spiral-like polarization structure. In RR~24b a strong 
velocity gradient in ionized gas could be associated with the polarization structure. We suggest that the  large-scale magnetic field of the RR~24 pair
members still  plays a role in shaping the polarization pattern. 
}
\maketitle

\section{Introduction}
 
During a galaxy-galaxy interaction, tidal forces strongly deform 
the gravitational potential affecting both the stellar and  gas distribution
and dynamics, which in turn are connected with the magnetic 
field structure of the galaxy \citep[see e.g.][and references therein]{Vallee97,Moss00, 
Widrow2002}.  
Recent studies, in the radio domain, suggest a possible connection between the
magnetic field structure and gas flows in spiral galaxies \citep[see e.g.][]{Beck99,Chyzy04}. 
These  authors suggest that the field is mostly frozen into the 
gas and follows its motion during galaxy encounters; contrary to the
expectation of a field generated by a galactic dynamo.

More recent studies have focused their attention on interacting  and 
merging spirals \citep[see e.g.][]{Soida06,Wezgowiec07,Vollmer07}.  
Interacting spiral galaxies indeed show both large
departures from a symmetric spiral shape and are expected to host gas
flows. Perturbed galaxies, then,  constitute a good laboratory for the study of the
interrelations between peculiar gas flows and magnetic field structure,
but no clear results emerge from the literature, mainly due to the
lack of high resolution kinematical information with which to compare
the magnetic field structure \citep[see e.g.][]{Soida01}. \citet{Chyzy04} 
found that the magnetic field in NGC~4038/4039, the Antennae,
is very different from what is observed in normal spirals. 
In particular, they notice that various regions in this merging 
galaxy reveal different physical  conditions and evolutionary stages. 
In particular, processes related to star formation tangle the
field lines, so that little polarization is observed in star-forming
regions. The magnetic field association with cold, warm and hot gas, 
depends on the particular place and dominance of various 
physical processes.  This leads to another open question, i.e. 
which gas phase has the strongest influence on the
magnetic field evolution.

Observations then, suggest that large-scale galactic magnetic fields 
evolve  during violent interaction episodes, which could lead
to accretion/merging phenomena. 

This study presents optical polarization maps of a sample of interacting galaxies. 
It is well known that interstellar dust grains get aligned with a magnetic field which
induces a polarization of the light passing through the dust cloud, via
a dichroic extinction. The study of the polarization has been 
used to gain information about the magnetic field in galaxies 
\citep[see e.g.][and reference therein]{Widrow2002,Lazarian07}.   
In particular, at optical and infrared wavelengths the analysis of 
 polarization has contributed to the mapping of large--scale magnetic field 
structure in spiral galaxies \citep[see e.g.][]{Scarrott96,Jones97,Jones00,Alton00}. 
Optical polarization maps reveal magnetic field structures, 
spatially coherent on scale-lengths of many kpc, both in {\it normal} 
and {\it active} galaxies although the presence of scattering phenomena 
makes the interpretation of the observations difficult \citep{Widrow2002}
in particular in dusty environments, where a significant amount of scattering
can occur.

\begin{table*}
\scriptsize{
\caption{\bf Overview of the sample}
\label{table1}
\begin{tabular}{lllllllc}
\hline\hline
            &  RR~23a    &  RR~23b      & RR~24a       &  RR~24b      &   RR~99a     &  RR 99b    & E2400100 \\
            &E1510361    & E1510360     &E2440120      &E2440121       &  E5520490    & E5520500    &            \\
            &            & NGC 454      &               &                 &   NGC 1738  & NGC 1739   &     a/b nuclei       \\
\hline
Morphol. Type                  &     Irr    &   Pec  & Sb[2] & S0-a[2] &  SB(s)bc pec: & SB(s)bc pec:& SAB0:pec  \\
Hel. Sys. Vel.   [km~s$^{-1}$] & 3626$\pm$2  &   3645 & 6852$\pm$24 [3]& 6719$\pm$55[3]& 3978$\pm$30 & 3982$\pm$32   &3159$\pm$20/3348$\pm$14  \\
                                                  &                         &                      & & & && \\
{\bf Apparent magnitude}     &                         &                     &  & & &&\\
{\bf and colours}:              &                         &                     &  & & &&\\
B$_T$                            &  13.11$\pm$0.09 &   13.12$\pm$0.21  &15.70$\pm$0.09 &14.43$\pm$0.09&  13.70$\pm$0.09  & 14.24   &12.60$\pm$0.30$$ \\
$\langle$(B-R)$_T \rangle$       &    1.02  &                            & 1.86 & 1.37   &   1.05  & 1.02      &1.39 \\
(J-H)$_{2MASS}$                    &             &                             &         & 0.708  &  0.574   & 0.551   &0.721\\
(H-K)$_{2MASS}$                   &            &                              &         &0.388   &    0.329 & 0.343   &0.195 \\
                                                 &            &                              &  & & & & \\
{\bf Galaxy structure}:                  &                             &             &  & & & &\\
Effective Surf. Bright. $\mu_e$(B) &    22.21$\pm$0.31   &   22.33$\pm$0.34  &             & 22.63$\pm$1.46 &  21.38$\pm$0.34  & 21.44$\pm$0.35     & 21.84$\pm$0.33\\
Average  P.A.                                &  57.3                      & 80.9                     &   131.9 & 31.6  &   35.6  & 104.4 &   132.5  \\
Average $\epsilon$                         &  0.26                      & 0.37                      &    0.49& 31.6 &  0.44    & 0.51  &   0.46 \\
                                                 &                          &                      &    & & &  &    \\
{\bf Kinematical parameters}        &                         &                       &    & && &\\
Vel.disp. $\sigma_0$ [km~s$^{-1}$]  stars   &         &    &  71$\pm$31& 80$\pm$66 &        &   &263$\pm$19/192$\pm$13 \\
Gas max. rotation V$_{max}$  [km~s$^{-1}$]      &                          &     &240 [3]& &205$\pm$9   &     &\\
                          &                          &                    &  & &    4          &        & 189\\
\hline
{\bf  Pair parameters}   &  RR~23    &      & RR~24      &     & RR~99    &   & E2400100 \\
\hline
Arp-Madore ident.   &     AM 0112-554       &  & AM0115-444&                      &    & &      \\
Projected separation [\arcmin]   &  0.45      &               & 0.3           &    &  0.53   &         & 0.08\\
$\Delta$ V  [km~s$^{-1}$]        & 19        &               & 133          &     &  4     &        & 189  \\
\hline
\end{tabular}

\medskip

\noindent{{\bf Note}-- The photometric and kinematical data are obtained from  {\tt NED} while 
structural data are from the {\tt HYPERLEDA} compilation. Detailed studies of RR~24 and 
ESO~2400100 have been performed by \cite{Rampa05} and \cite{L98b}. 
In particular, ESO 2400100,  is considered a single early-type galaxy 
showing a shell system but it is, rather, composed of two distinct components, indicated 
in the table with $a$ and $b$ \citep{L98b,Rampazzo03}.}}
\end{table*}

With this pilot study we aim to investigate the ability of  optical
polarization maps to trace the evolution of the large--scale 
magnetic field structure during different {\it phases} of the interaction
phenomenon. Furthermore, polarization maps may unveil 
important effects induced during galaxy encounters, 
such as the displacements of dust from its more usual stellar-disk
enviroment \cite[see e.g.][]{Alton00}. We choose to investigate
pairs of galaxies in very low density environments. 
\citet[][]{Wezgowiec07} have shown that the cluster environment 
may affect the large-scale magnetic field, not only
by galaxy interactions, but also by the interaction
between the Inter Galactic Medium and the Inter Stellar 
Medium,  e.g. due to ram-pressure.

The paper is organized as follows. Section~2  presents 
the sample with the aim of characterizing the interaction episode 
according to the  morphological and kinematical  effects induced on 
each member galaxy. Section~3 details the observations and the reduction techniques. 
Results are organized as individual notes on paired galaxies in Section~4. 
In Section~5, we discuss the polarization maps in the context of 
the kinematics of the galaxies by comparing the polarimetric maps 
with  the 2D gas  velocity fields. We 
investigate whether the polarization vectors follow gas flows or 
maintain/develop their own structures independently. Conclusions and
future prospects are sketched in Section~6.

\section{The sample}

We consider a sample of four interacting pairs of galaxies.  Three of them
are selected from the \citet{rr95} catalogue of pairs  in the Southern 
Hemisphere. Of the three pairs, two, namely RR~24 and RR~99, are 
composed of spiral members. Both members in RR~23 are of uncertain
morphological classification (see Table~\ref{table1}).
 RR~23a, the western member, is classified 
as Irr but it is probably  a distorted spiral. The companion, RR~23b,
is classified as Pec, although  \citet{Stiavelli98} and \citet{PRR00}
suggest that it is a typical S0. RR~23 is then a mixed morphology pair. 

ESO~2400100 has two well--defined nuclei embedded in a single envelope. 
It is indeed classified as a single early-type galaxy showing a large shell
system \citep{mc83}. According to simulations, shell galaxies represent 
the debris of an accretion/merging event. 

The average velocity separation of the four pair members is 
$\langle\Delta~V\rangle$ = 84$\pm$76 km~s$^{-1}$, suggesting that 
they form physical entities. The highest velocity separation, 189 km~s$^{-1}$,
is shown by the two nuclei of ESO~2400100$a$ and $b$. \citet{L98b}
have shown that the two nuclei are gravitationally interacting.  

The distorted morphology of member galaxies in some pairs, in particular
RR~23 and RR~24, suggests that the stars and the gas are already strongly 
influenced by tidal forces.  In these cases simulations predict the presence of 
non--radial motion, gas flows  and stripped tails of gas. 
\citet{Rampa05} provided direct  evidence of such gas flows and tidal  
streaming in the 2D  H$\alpha$  velocity field of RR~24 and
ESO~2400100.    

We summarize the basic data for the pairs  in Table~\ref{table1}.\\

\section{Observations and data reduction}

\subsection{Observations}

Observations were carried out in 2004 November at the 3.6m ESO telescope 
(La Silla, Chile) using  EFOSC-2 (ESO Faint Object Spectrograph and Camera) 
used in the polarimetry mode \citep{Hau2003}.  The CCD
used was the Loral 2K back thinned illuminated, ESO\#40.  The seeing
conditions were good, with values of FWHM ranging from 0\farcs8 to 1\farcs2.

EFOSC-2 is a classical dual beam polarimeter with a
halfwave plate which permits the rotation of the state of the polarization of
the light entering the telescope. A sequence of four images with 
four different positions of the half wave plate (0$^\circ$, 22.5$^\circ$,
45$^\circ$, 67.5$^\circ$) was  taken. Rotating the plate the
polarization state of  each channel is thus  0$^\circ$,
45$^\circ$, 90$^\circ$ and 135$^\circ$. Because of the presence 
of the polarimetric mask in the focal plane, the sequence has to be 
repeated after an offset of the telescope of 20\arcsec\ (the pitch of 
the polarimetric mask) to cover homogeneously the field of view.  
Each observation  in the R band was completed during the same night,
as detailed in Table~\ref{table2}, with a batch procedure.

Dome flat fields were taken with the halfwave plate continously rotating 
during the acquisition. Polarimetry observations were calibrated 
through the acquisition of  standard unpolarized (WD-2359-434, 
WD-0752-676) and polarized stars (NGC 2024) each night. The EFOSC-2 
polarimeter presents a residual instrumental polarization well below 
0.1\%.

\subsection{Data reduction}

Bias subtraction, flat fielding and removal of the cosmic ray signatures 
were performed using standard {\tt IRAF}\footnote{{\tt IRAF} is distributed by 
the National Optical Astronomy Observatories, which are operated by the 
Association of Universities for Research in Astronomy Inc., under cooperative 
agreement with the National Science Foundation.} packages.

After a careful sky subtraction and fine alignment of the images obtained
from different frames and  channels, we finally obtained  images 
of the linear components of the Stokes parameters $(Q, U)$ and 
the polarized intensity $P=(Q^2+U^2)^{1/2}$. 
The $Q$ and $U$ quantities are defined as:

$$ Q  = \frac{I_{0}-I_{90}}{I_{0}+I_{90}} $$ $$ U  =
\frac{I_{45}-I_{135}}{I_{45}+I_{135}} $$

where the couples ($I_{0}$, $I_{90}$) and   ($I_{45}$, $I_{135}$) 
are the photon counts for each pixel in the corresponding channel.
The Stokes parameters are calculated pixel by pixel, through an 
image rebinning, when necessary, in order to reach an 
adequate S/N. The final calibration is obtained through
a vectorial subtraction from the target polarization $(Q, U)$ of the
residual instrumental polarization $(Q_0, U_0)$ obtained from
measurements of an un-polarized standard star. 

The  error in polarization  is
$$\Delta~P =  {{1} \over {P}} \sqrt{(Q \Delta Q)^2 + (U \Delta U)^2} $$ 
while the error in polarization angle can be approximated by
$$\Delta\theta \simeq {{1} \over {2}} arctan (\Delta~P/P)$$
\citep[see e.g.][]{Fendt98}.

The entire reduction process was performed through a specific in house
software package developed making use of the commercial {\tt IDL} 
language and of the {\tt IDL Astronomy User's Library} (Landsman 1995).

\begin{table}
\caption{\bf Details of observations}
\begin{tabular}{lc}
& \\
\multicolumn{2}{c}{} \\
\hline\hline
\multicolumn{1}{l}{ident.}&
\multicolumn{1}{c}{total exposure time}  \\
\multicolumn{1}{c}{}& 
\multicolumn{1}{c}{[sec]} \\
\hline
RR~23a  & 8$\times$600 \\
RR~23b & 16$\times$100 \\
RR 24a  & 16$\times$512 \\
RR 24b  & 16$\times$300 \\
RR 99a  & 8$\times$600 \\
RR~99b & 8$\times$600 \\
ESO 2400100 & 16$\times$150 \\
\hline
\end{tabular}
\label{table2}
\end{table}

\begin{figure}
\begin{center}
\includegraphics[width=0.5\textwidth]{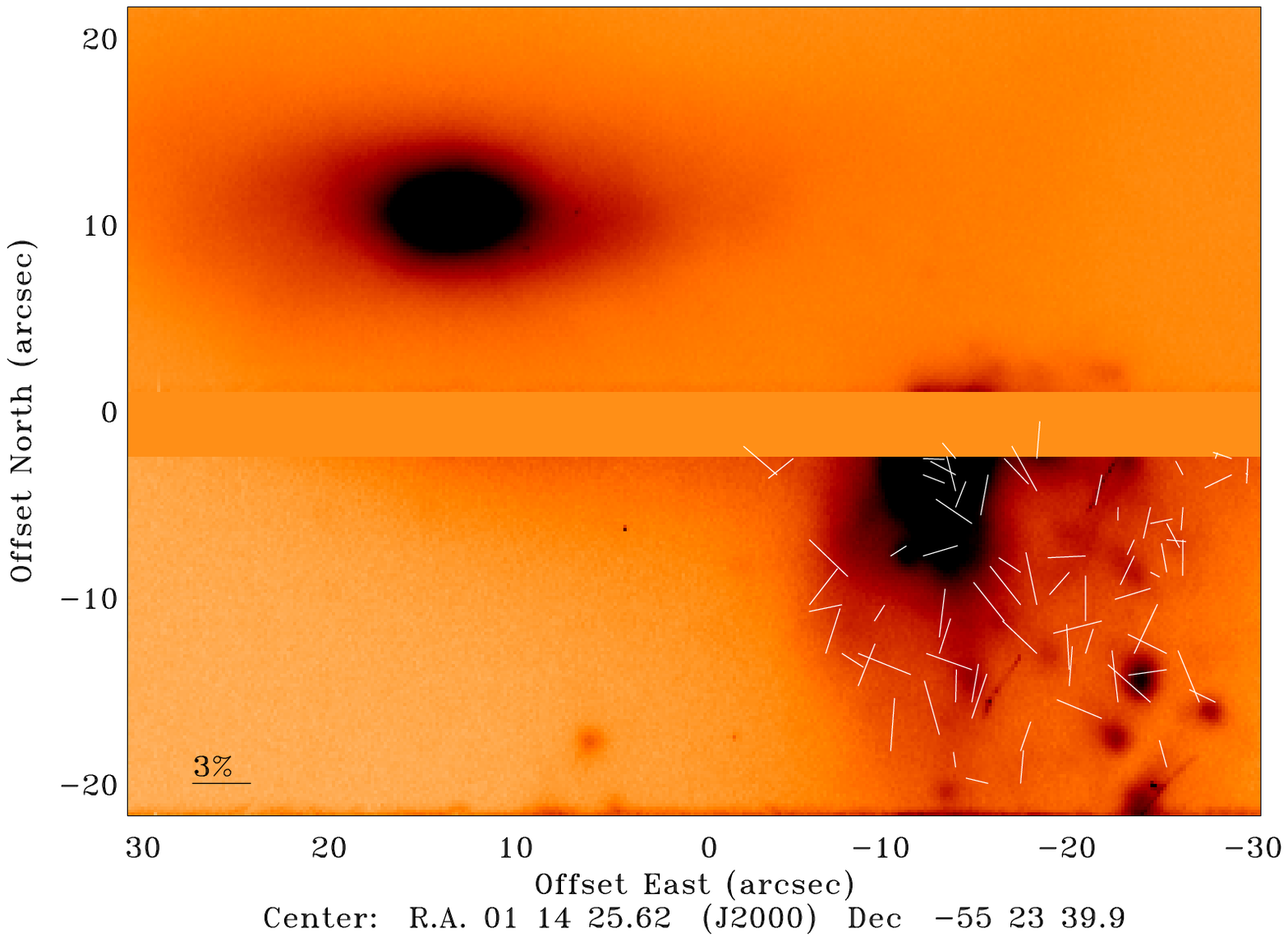}
\includegraphics[width=0.5\textwidth]{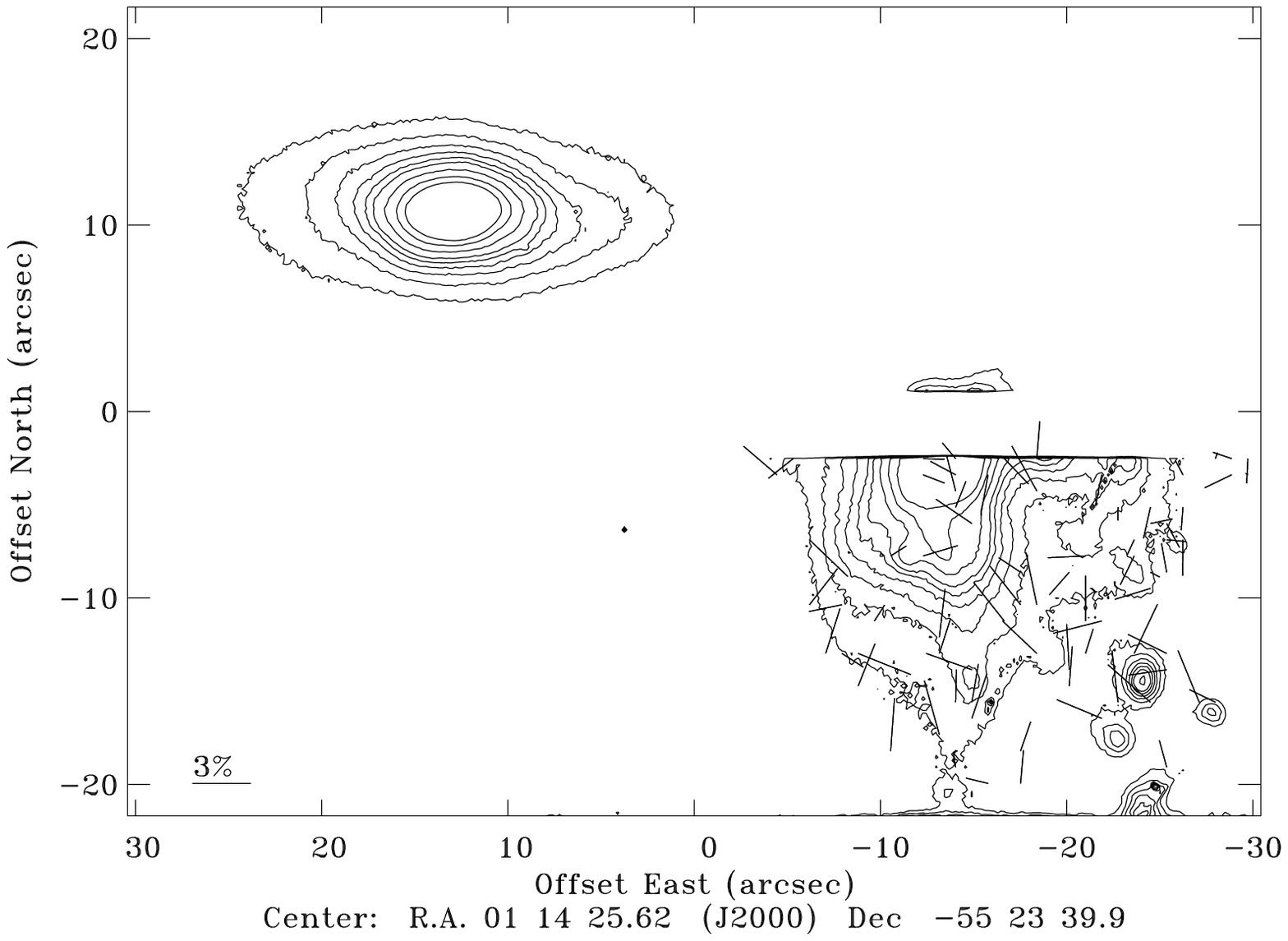}
\caption{Linear polarization vectors overlayed on the 
R-band image (top panel) of  RR~23a and RR~23b, 
the southern and the northern member, respectively, 
and on the galaxy contours (bottom panel). 
The length of the lines is proportional to
the polarization degree according to the scale in the lower left corner
of the plot.  We clip the maps at a polarization level
0.1\% $\leq P \leq$ 3\% and  a polarization error, 
$\Delta~P < 1P$. The error in the polarization angle, $\Delta\theta$,
is $<22^\circ$.} 
\label{fig1}
\end{center}
\end{figure}

\section{Results}

In this section we present our linear polarization measures.
For each pair we plot the linear polarization maps,
superimposed on the R-band image and the intensity
contours of the pair.
The data are displayed in the conventional manner, 
with vectors proportional to percentage of the linear 
polarization. 
We plot polarization vectors within selected limits 
$P_{min} \leq P \leq P_{max}$ which we indicate
in figure captions together with the polarization error, $\Delta P$,
and the error in the polarization angle, $\Delta\theta$.

We also considered the maps of the polarized intensity $(Q^2+U^2)$,
bias corrected, to trace the signal in those parts of the galaxy
where the degree of polarization is low, i.e. vectors are so short that
no orientation is visible \citep[see e.g.][]{Fendt98}. The bias correction
has been performed following the recipe in \citet[][see Appendix]{Wardle74}. No significant
signals/structures are detected in the polarized intensity maps of the pairs, 
except in the RR~24 pair as discussed in the following paragraphs.

The optical polarization due to the dichroic  extinction, caused
by grains partially aligned  with a magnetic field, 
is parallel to the orientation of the 
field \citep[see e.g.][]{Purcell79}. A spiral polarization
pattern in face-on galaxies may be interpreted  as caused by 
an underlying magnetic field according to radio observations.
Notice, however, that in some cases the vector pattern does not 
follow the spiral arms. Relevant are the cases of the pair in M~51 
\citep{Scarrott87}  and NGC~6946 \citep{Fendt98}. 
These deviations from the spiral pattern are observed typically 
in the outer part of the galaxies, where the polarization 
structure could be influenced by effects external to the galaxy.
Scattered light is polarized perpendicular to the 
scattering plane, which is defined by the propagation vector of the incoming and 
outgoing light ray. A circular pattern is a clear indication of anisotropic scattering,
in particular if  there is a bright nucleus as scattering source 
\citep[see e.g.][and reference therein]{Alton00}. At the same time,
there are situations in which the origin of the polarization is of much more
difficult interpretation \citep[see the case of NGC~891 in][]{Fendt96}. 

Individual results are preceded by notes, taken from the recent literature
describing the pair properties with the aim of characterizing the phase of
the encounter.

\noindent
\underbar{RR~23} ~~~~~ RR~23a, the southern member (Figure~\ref{fig1}), is severely disrupted and a complex system 
of debris extends from the galaxy towards the SW. No spiral arm
structure is recognizable. The companion, RR~23b,  is a typical S0 
galaxy with a central dust-lane  \citep{Stiavelli98,PRR00}. 
RR~23b is warped in the outer parts and
\citet{Stiavelli98}  suggest that a substantial amount of gas has sunk 
into the center of this component.  
\cite{Tanvuia03} measured the presence of an ionized 
gas component with a H$\alpha$/[NII]$\lambda$6583 line ratio
consistent with the presence of nuclear activity, although no 
[OIII]$\lambda$5007 emission has been detected. 
\citet{Stiavelli98} find neither star formation activity nor nuclear activity. 

In Figure~\ref{fig1} we show the linear polarization map of the pair members
and we provide in the caption the error in the  polarization estimate.
There is no significant polarization in either galaxy, even in the central 
nuclei, nor in the bright knots visible in RR~23a.  

The surface photometry of \citet{Couto06} clearly shows  the presence of dust in their residual map (see their Figure 7) of RR~23b but its contribution to the extinction could be low.  In the case of RR~23a the lack of polarization 
could be connected ether with the presence of wide, young 
star forming regions detected by \citet{Stiavelli98} \citep[see also][] {Chyzy04}  
or with the high degree of the galaxy distortion.
 
\begin{figure*}
{\hspace{-0.5cm}{
\includegraphics[width=12.5cm]{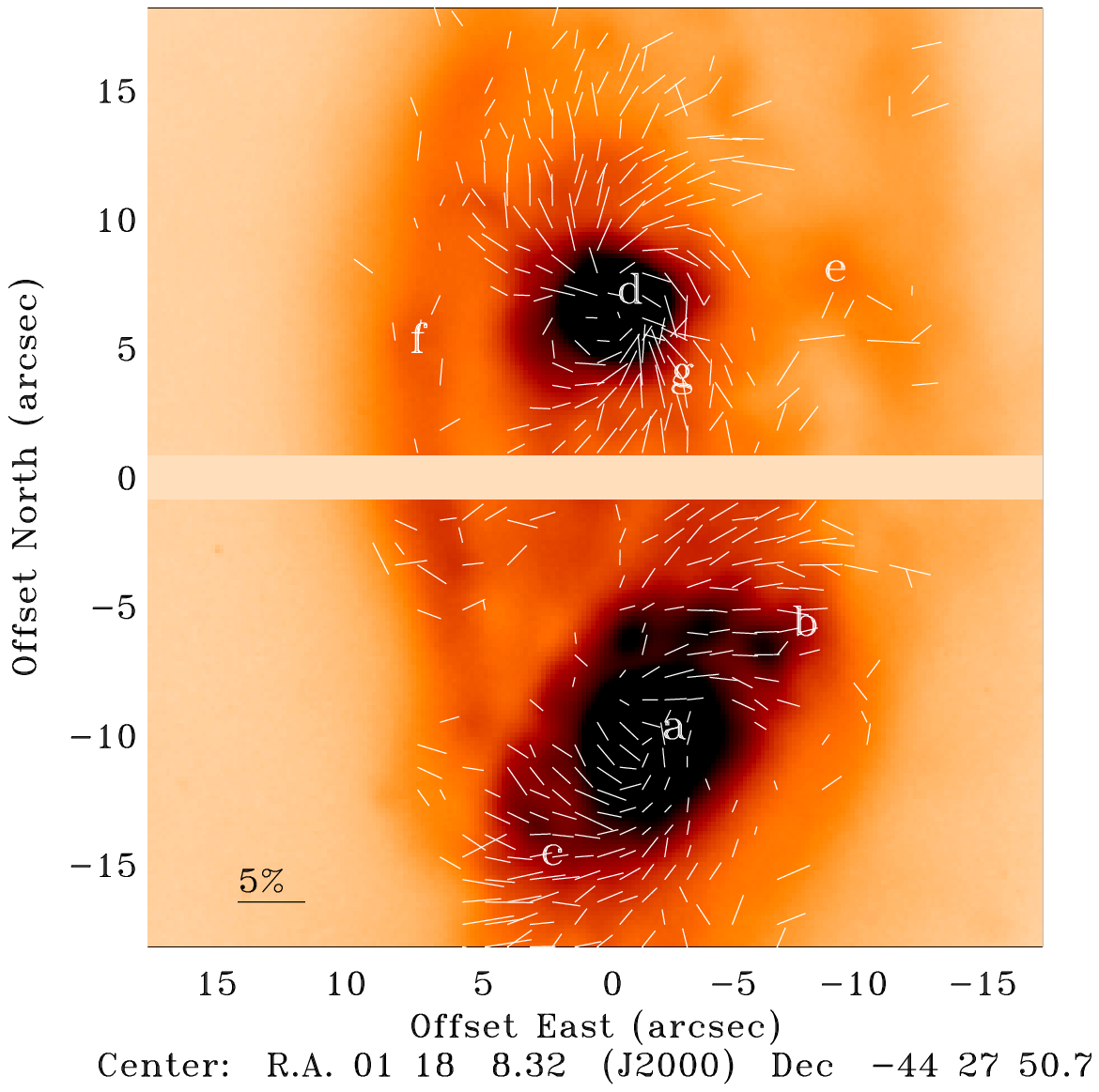}
\hspace{-4cm}
\includegraphics[width=12.5cm]{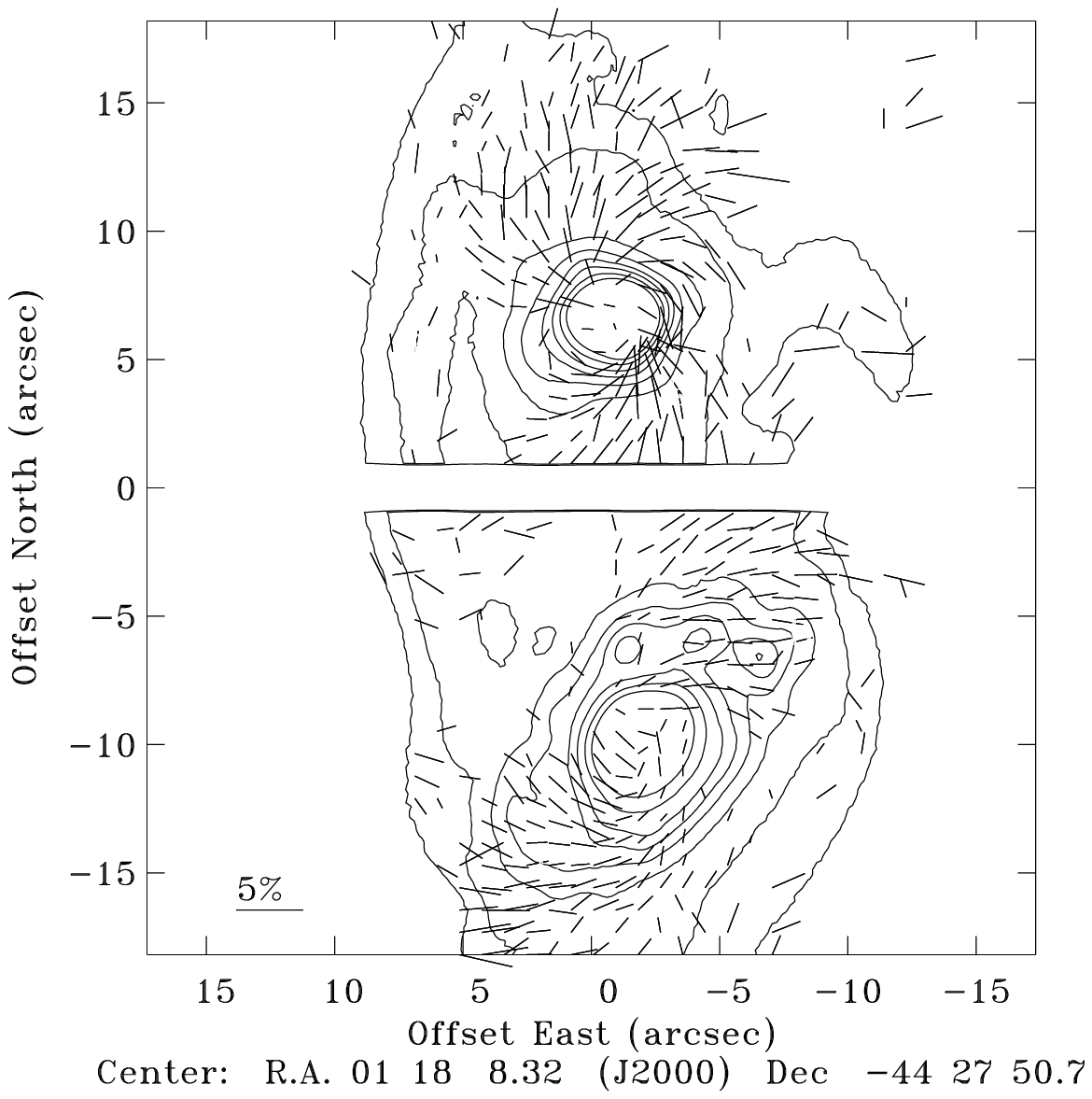}}

\vskip0.7cm
\includegraphics[width=12.5cm]{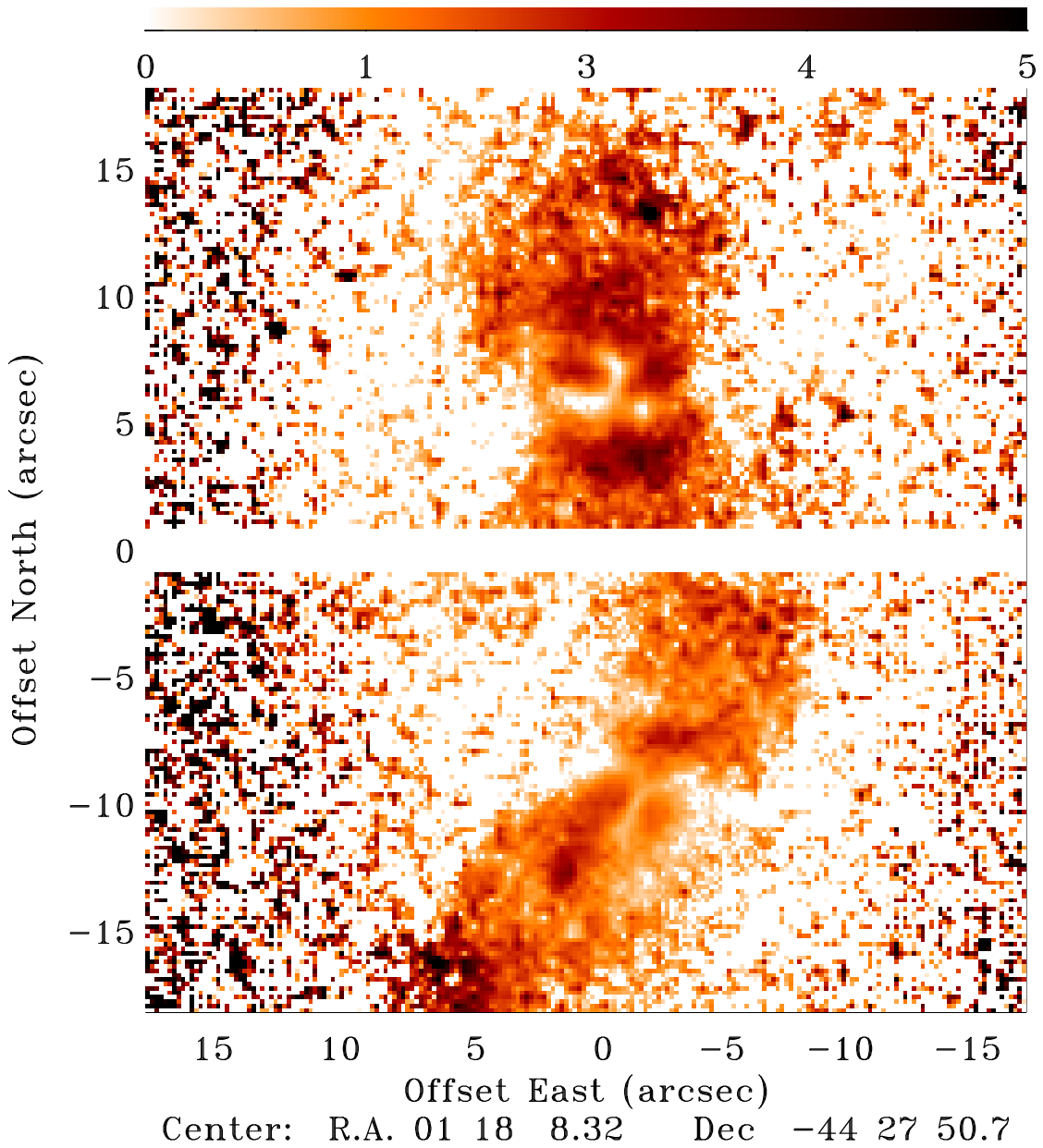}
}
\caption{ Linear polarization vectors superposed upon the 
R-band image (top left panel) of  RR~24a, the southern member,
and RR~24b.  In the top right panel polarization vectors 
are superposed upon the galaxies intensity contours. In the bottom panel
we plot the polarized intensity $(Q^2+U^2)$, bias subtracted,
image. We clip the maps at a polarization levels
0.1\% $\leq P \leq$ 5\% and  at a polarization error 
$\Delta P < 0.8P$. The error in the polarization angle, $\Delta\theta$,
is $<19^\circ$.}
\label{fig2}
\end{figure*}

\begin{figure*}
\includegraphics[width=12.5cm]{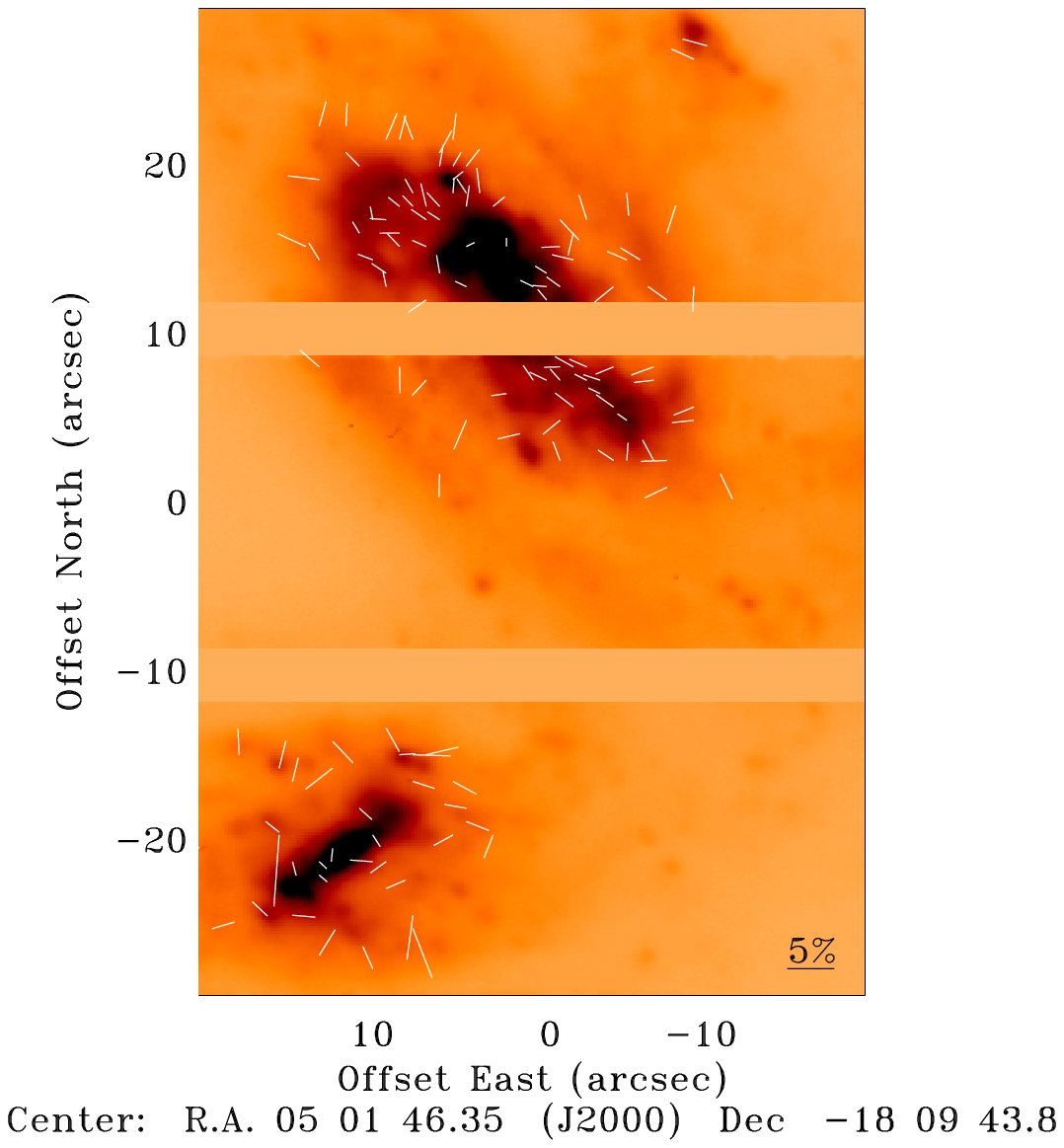}
\hspace{-4cm}
\includegraphics[width=12.5cm]{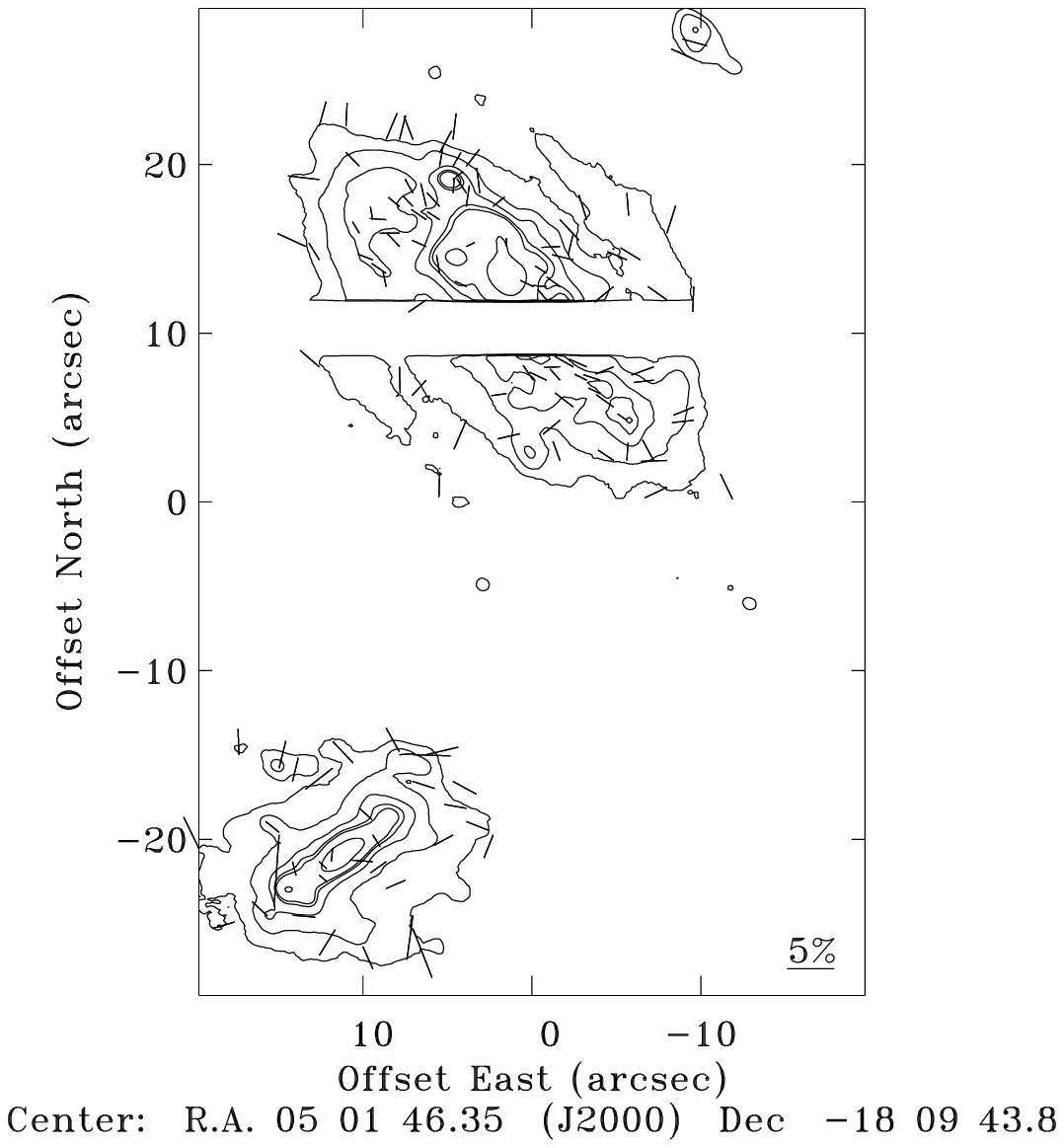}
\caption{ Linear polarization vectors overlayed on the 
R-band image (top panel) and contours (bottom panel) 
of RR~99. RR~99a is the northern galaxy. 
We clip the maps at a polarization level
0.1\% $\leq P \leq$ 5\% and  a polarization error, 
$\Delta P < 1P$. The error in the polarization angle, $\Delta\theta$,
is $<22^\circ$.} 
\label{fig3}
\end{figure*}

\noindent
\underbar{RR~24} ~~~~~   The  pair, also known as VV~827
\citep{vv77} and AM0115-444 \citep{am87}, is shown in Figure~\ref{fig2}. 
It is a bright IR emitter (L$_{FIR}$=1.2 $\times$ 10$^{11}$ L$_\odot$). 
The 2D(B-V) color of the two objects is quite similar and consistent with the
late--type morphological classification \citep{rr96}. The distorted structure of
the arms, in particular in the northern member RR~24b, suggests that they
are probably created/modified by the interaction. 
\citet{L98b} and \citet{agu00} studied the stellar and gas kinematics of
the RR~24 system. They measured a systemic velocity separation of 
$\approx$189 km~s$^{-1}$ and showed that the position--velocity profile 
of the ionized gas, along the line connecting the galaxy nuclei,  is consistent
with that of the stars. They further suggested that the U-shape of the
rotation curve could reflect the ongoing interaction during an
interpenetrating encounter of the two galaxies \citep{comb95}. 
\citet{agu00} report  that the
equivalent with of the H$\alpha$+[NII] lines of the RR~24b nucleus,
compared with isolated galaxies, strongly reflects an important
star formation activity, indicating the presence of a starburst, probably induced by the 
interaction. The equivalent width of RR~24a is compatible with a ``normal'' Sb
galaxy. They conclude that around the nucelus of RR~24b, the oxygen and
nitrogen, the excitations, the equivalent width, and the internal
absorption are systematically different from those of the southern 
nucleus, indicating that they have followed different evolutionary path.

The linear polarization map and polarized intensity plot  are shown 
in Figure~\ref{fig2}. Average polarization intensity and the average
polarization angle values have been mapped in several regions 
within the pair and are reported in  Table~\ref{table3}. 
Values have  been calculated within  square boxes of  
2"$\times$2" (i.e. $\approx$0.9 kpc,
H$_0$=75 km~s$^{-1}$Mpc$^{-1}$).
The $a$ and $d$ regions correspond to the RR~24a and RR~24b nuclei. The regions
$b$ and $c$ are selected on the northern and southern arms of RR~23a, respectively.
Regions $e$ and $f$ lie in the areas of RR~24b where the 2D H$\alpha$ kinematics 
shows the presence of tails of gas falling onto RR~24a (see Figure~\ref{fig5} (left panel)
adapted from \citet{Rampa05}). At position $g$, we intend to map an intense linearly
polarized area in RR~24b. In $g$ the linear  polarization reaches a peak of 
3\%$\pm$0.7 while in the other mapped areas it is 1-2\%.

\begin{table}
\caption{\bf Summary of  results in RR~24}
\begin{tabular}{llcc}
\hline\hline
\multicolumn{1}{l}{member}&
\multicolumn{1}{l}{area}&
\multicolumn{1}{c}{average}  &
\multicolumn{1}{c}{$\theta$}\\
\multicolumn{1}{l}{}&
\multicolumn{1}{l}{}   &
\multicolumn{1}{c}{Polariz.(\%)}  &
\multicolumn{1}{c}{NE}  \\
\hline
RR 24a          &  (a) S nucleus  &  1.25$\pm$0.32  & 53.8$\pm$6.6 \\
                     &  (b) North-West arm   &  1.19$\pm$0.70  &  95.6$\pm$15.0 \\
                     &  (c) South-East arm                &  1.47$\pm$0.86  &  87.7$\pm$15.3 \\
RR 24b         &  (d) N nucleus &  1.15$\pm$0.23  &  84.4$\pm$5.2\\
                     &  (e) North-West tail              &  0.54$\pm$1.81  & 87.6$\pm$31.0\\
                     &  (f) East tail                          &  0.67$\pm$1.04  & 112.6$\pm$ 22.5\\
                    &   (g) between galaxies          &  3.26$\pm$0.75  &  53.2$\pm$ 6.6\\
\hline 
\end{tabular}
\label{table3}
\end{table}

The polarization map indicates that the two galaxies 
have coherent polarization structures that extend for several kiloparsecs. 

In RR~24a, the southern member of the pair, the polarization vectors 
follow the spiral structure, both in the arm and in the inter-arm
regions. The polarization intensity map, Figure~\ref{fig2} bottom panel,
shows a "butterfly-like" pattern. 
In RR~24a there are extended polarized areas north
and south of the arms indicated with $b$ and $e$. 
We interpret the above polarization structure  as
due to dichroism by analogy to several nearby spiral galaxies 
studied in the radio and optical domain. We suggest that
 the polarization vectors follow the configuration of the
large-scale magnetic field in the galaxy.

Unlike the companion, the morphology of RR~24b is very
complex. The very long outer arms/tails, labeled $f$ and $e$,
have probably been generated by stripping during the ongoing interaction
(see also Figure~\ref{fig5} and discussion next Section).
This interaction episode is so violent that it has also probably modified 
the inner part of the galaxy. 
A ``S-shape" polarization pattern departs from the galaxy center 
towards the South-West ($g$) and towards the  North, 
seems to trace the presence of open, more diffuse arms. The
Northern polarization pattern seems uncorrelated  with
the  filamentary arms/tail indicated with  $f$ and $e$ and may
trace the pristine spiral structure of the galaxy. Open arms
are often generated in disk galaxies during the first phase of an interaction 
episode as shown by simulations \citep[see e.g.][]{Nogu88}.
The high polarization value measured in $g$ possibly maps 
a displacement of dust from the galactic plane.

Due to the galaxy distortion the  inclination of RR~24b  is unclear,
hindering considerations about polarization 
effects due to scattered light, although it could be present.
Figure~\ref{fig2} shows that the bright central core is
the dominant light source in this galaxy and that it
is not the center of  the polarization vector pattern, as might be
expected if it were the scattering source  \citep[see e.g. ][]{Scarrott96,Alton00}.
Instead, the central core, $d$, is crossed by  the polarization vectors
with a pattern that is reminiscent of open arms as described above.
In this context, we suggest that the  large-scale magnetic field of the galaxy still  plays a role in shaping the polarization pattern.

\noindent
\underbar{RR~99}~~~~~ The pair, shown in Figure~\ref{fig3},
is composed of two late-type galaxies \citep[see e.g.][]{rr96}. 
\citet{White2000}, from their B and I band imaging, noticed
the strongly asymmetric structure of this Sbc pair system due to the
ongoing interaction. Both galaxies show multiple arms 
and HII regions  irregularly distributed within the plane
of the galaxies. 

In Figure~\ref{fig3} we show the linear polarization map.  
We notice a possible polarization pattern in RR~99a with
vectors aligned along the plane of the galaxy. 
\citet{White2000} measured a quite low
 face-on extinction in the region between the two galaxies  
(see their Figure 4): in the arm region, $A_B$=0.3-0.37 
and $A_I$=0.24-0.28, while in the inter-arm region,  
$A_B$=0.2-0.26 and $A_I$=0.17.

\begin{figure}
\centerline{\includegraphics[width=0.5\textwidth]{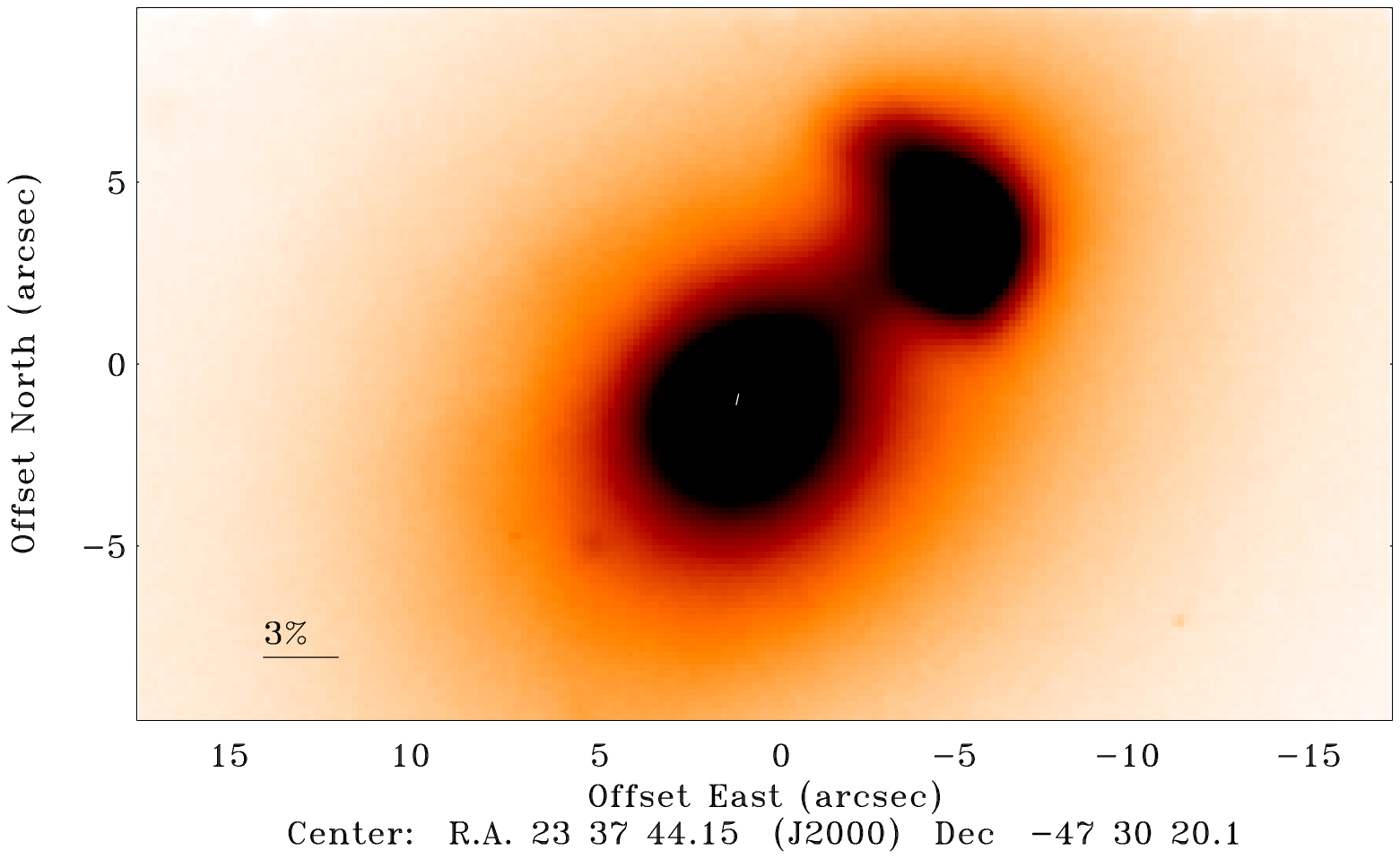}}
\centerline{\includegraphics[width=0.5\textwidth]{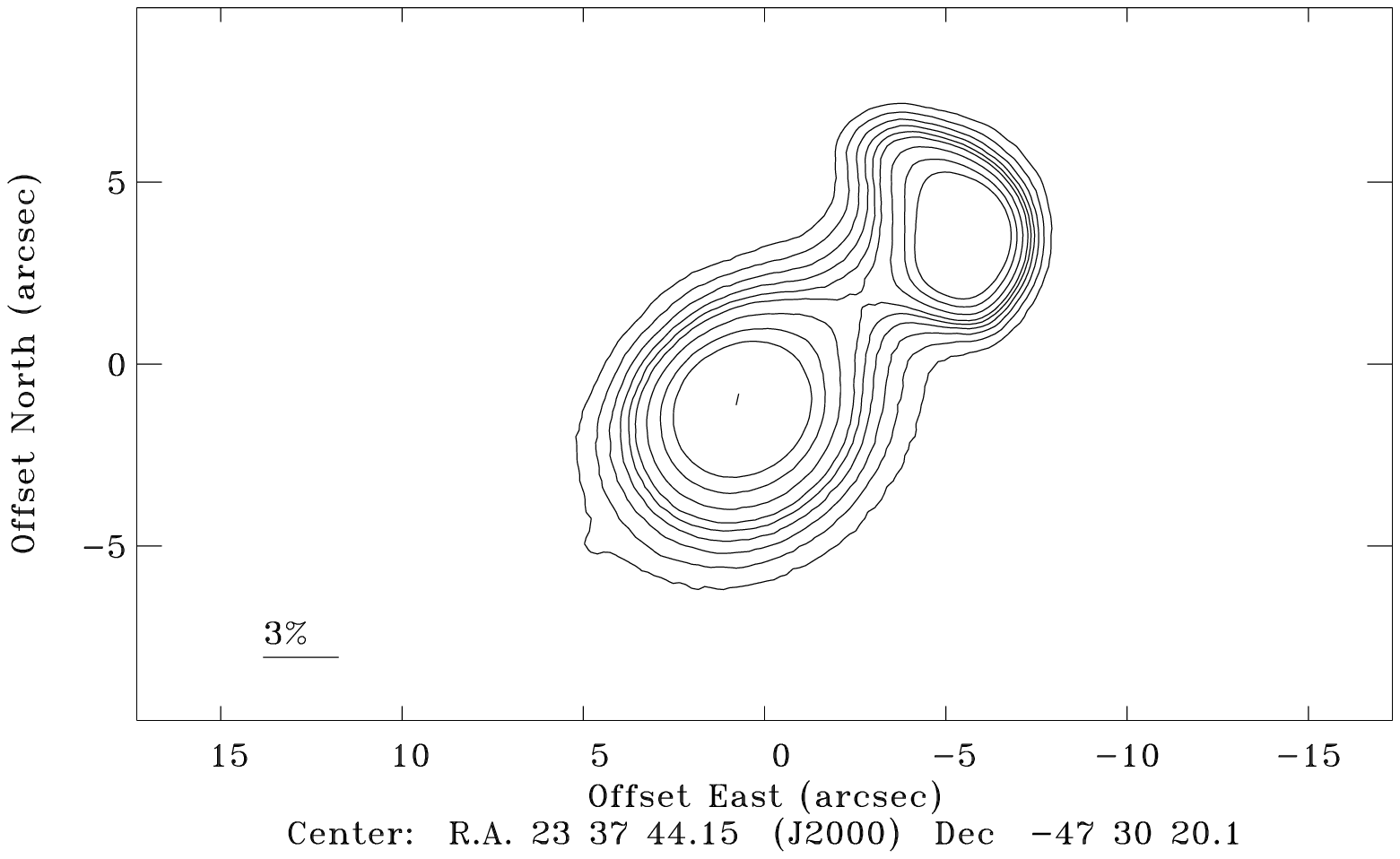}}
\caption{ Linear polarization vectors overlayed on the 
R-band image (top panel) and contours (bottom panel)
of ESO~2400100. The system is catalogued as a 
unique early-type galaxy with shell  in \citet{mc83}. ESO~2400100a in
Table~\ref{table1} is the northern nucleus, ESO~2400100b the southern
\citep{L98b}.  We clip the maps at a polarization level
0.1\% $\leq P \leq$ 5\% and  a polarization error, 
$\Delta P < 1P$. The error in the polarization angle, $\Delta\theta$,
is $<22^\circ$.} 
\label{fig4}
\end{figure}

\begin{figure*}
\resizebox{17cm}{!}{
\includegraphics[]{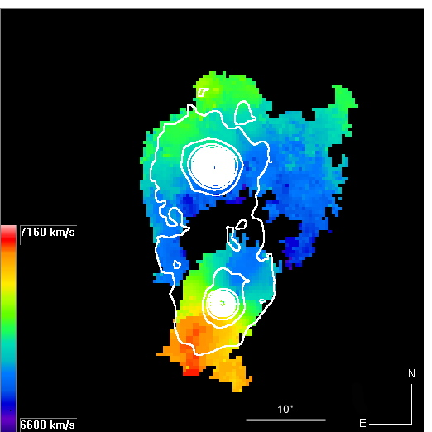}
\includegraphics[]{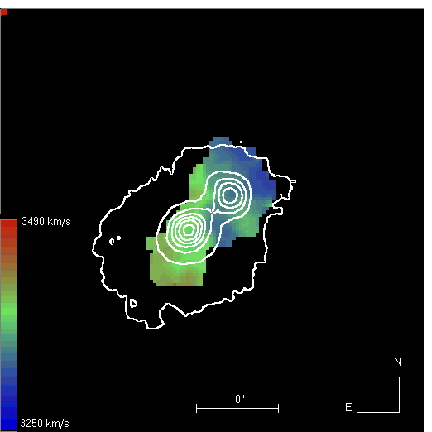}
}
\caption{ The two panels are adapted from \citet{Rampa05} (left panel) and
\citet{Rampazzo03} (right panel). The white 
contours trace the shape of the stellar isopothes from the continuum 
nearby H$\alpha$.  (left panel)  Velocity field of the RR~24 system. 
The warm gas velocity was measured in several zones  along the east 
and the west arm/tails of  RR~24b. The lines profiles show a continuity 
in the velocity indicating the presence of a gas flow along the tidal tail.
(right panel) Velocity field within the shell galaxy ESO~2400100. The velocity
 field is irregular and extends over the two nuclei.} 
\label{fig5}
\end{figure*}

\noindent
\underbar{ESO2400100}~~~~~ Studying the kinematics of this object,
\citet{L98b}  found two distinct nuclei separated by
$\approx$ 5\arcsec and $\approx$200 km~s$^{-1}$ which had not been identified in
previous papers \citep{mc83,cart88}. The velocity profile which \citet{L98b} 
measured along the line connecting the two nuclei shows a peculiar U-shape.
Simulations of interpenetrating encounters \citep[][and
references therein]{comb95} show that the U-shape, and the correlated 
inverted U-shape, in a velocity profile obtained along the line connecting 
two galaxy nuclei is a {\it bona fide} signature of their ongoing 
interaction, although this peculiar profile develops during both 
bound and unbound encounters.  In this framework,
ESO 2400100 could represent either a rare snapshot of shell formation, 
or an encounter  between two galaxies, one of which has
already suffered a recent accretion episode which has
generated the shell system \citep{Dupraz86,Thomson91,Trinchieri08}. 

Figure~\ref{fig4}  displays only the central part  of
ESO~2400100. This early-type galaxy shows a large system of shells, mainly
visible in its outskirts. ESO~2400100 has an extended ionized gas distribution connecting the two nuclei \citep[see][Figure~\ref{fig5}]{Rampazzo03}. Such gas is usually associated with dust in early-type galaxies \citep[see e.g.][]{Gou98}. 

We overlay to the image in Figure~\ref{fig4} the linear polarization 
vectors showing that the two nuclei are not polarized. 

\section{The linear polarization map and the 2D velocity field}

Simulations show that interaction and accretion events create new structures in 
the stellar disk, like bar and rings, and can
form merging remnants that resemble an elliptical galaxy
\citep[see e.g.][]{Barnes96,Schweizer96}. 
Encounters of gaseous disks  are expected to give rise to gas flows, 
galaxy-scale shocks and velocity gradients 
\citep[see e.g.][]{Nogu88,Gerin90,Salo91,Miwa98,Berentzen04}. 

Large--scale magnetic fields could be reconfigured by local
motions of the interstellar gas. This makes spiral galaxies in
interacting systems key  targets for connecting the properties of the
Inter Stellar Medium phases, different physical processes and their magnetic field
structure. At odds with  the vast majority of normal spiral galaxies, 
with regular spiral-like magnetic fields closely associated with the material spiral arms 
\citep[see e.g.][]{Beck1996}, disrupted, interacting spirals show strong 
departures from symmetric gas flows.
Magnetic fields trace regions of gas compression, strong shear and enhanced 
turbulence \citep[see e.g.][] {Chyzy04,Chyzy08}. In  interaction generated bars,  
the magnetic field could be aligned with gas streamlines along the bar
\citep[see e.g.][]{Beck99,Moss07}.

It has thus often been stressed that the 
velocity field of the gas should be measured
in detail to provide input for understanding the origin and evolution of
magnetic fields. For two pairs in the present study, namely  RR~24 and 
ESO 2400100,  the gaseous  and stellar velocity fields are known.
The Fabry-Perot study of \citet{Rampa05}  (Figure~\ref{fig5} left panel)
indicates that in RR~24 members are connected by a tidal tail of  ionized gas, 
through which the kinematically disturbed component RR~24b seems to supply gas to
RR~24a.  The gas velocity field in RR~24a is substantially unperturbed, while
that of the northern companion is completely distorted by the ongoing
interaction. The 2D kinematical study of ESO~2400100 (Figure~\ref{fig5} right panel;
see \citet{Rampazzo03})  shows a shallow velocity gradient
between the  two  H$\alpha$ peaks. These are displaced in opposite directions 
roughly perpendicular to the major axis defined by the outer gas envelope 
encircling the two nuclei. The phenomenon 
is expected during interpenetrating encounters \citep{Davoust88,comb95}.  

The regular velocity field of the ionized gas
measured by \citet{Rampa05} in RR~24a argues in favor of a limited 
perturbation of the galaxy induced by the companion. This suggests that the 
spiral--like shape of the linear polarization measured in RR~24a likely arises 
from dichroic extinction by magnetically aligned dust grains, as in
unperturbed galaxies \citep[see e.g.][]{Scarrott96}. 
Observation would then suggests that the magnetic 
field in this galaxy still has an ordered configuration, notwithstanding 
the ongoing interaction. 

In the case of RR~24b the morphological perturbation 
reflects the ionized gas velocity field, 
which is completely irregular and presents tails of gas falling onto the 
companion. At the same time, the ``S-shape" polarization pattern, shown in 
Figure~\ref{fig2}, corresponds to the strongest velocity gradient  in the ionized
 gas kinematics (see Figure~\ref{fig5} in the North-East,
   South-west direction crossing the nucleus). 
Although, the 2D kinematics  depart from the normal rotation pattern
(see e.g. the companion), we suggest that this is the main body of the galaxy 
where the  large-scale magnetic field is still  playing a role in shaping the 
polarization pattern.

Concerning ESO~2400100, it is known that the optical polarization 
is of limited value as a probe 
of the extragalactic magnetic fields, since dust is the tracer of the 
polarization: the effect depends on extinction. 
There is approximately one magnitude of visual extinction for each
3\% of polarization \citep{Scarrott87}. The amount of dust could be
small  with respect to a ``mature'' shell galaxy. This is also the case for
RR~23b, the early type member of the pair.
In addition, the kinematics of the gas and the stars in the two nuclei
in ESO~2400100 are completely decoupled, probably due to 
the very advanced phase of the encounter in ESO~2400100
with respect to RR~24. 

In our view, a 2D kinematical investigation
of the RR~23  and the RR~99 pairs should provide significant  information 
about the phase and the degree  of galaxy disruption. The velocity field
could show that both RR~99 members and RR~23a are too disturbed to 
preserve a pattern in polarization (and possibly in the magnetic field) 
as in the case of RR~24.

\section{Conclusions}

We present a linear polarization study of a pilot sample of interacting 
galaxies, consisting of three pairs containing late-type members
and a double nucleus shell galaxy. The latter should represent the 
fossil evidence of an accretion event, i.e. the final phase of 
a bound galaxy encounter.   

We find the following:
\begin{itemize}

\item{We did not detect a significant polarization in the very distorted late-type members, RR~23a, RR~99a and RR~99b. It is unclear
if the lack of a polarization pattern is connected to the high degree of galaxy 
disruption.} 

\item{The two early-type galaxies RR~23b and ESO~2400100  do not show any  
linear polarization. The lack of polarization may be due
to the low gas and dust content.
Early-type galaxies may lack gas and Inter Stellar Medium
 turbulence, hence cannot drive a dynamo and may not host large-scale magnetic fields.}   

\item{Only the pair RR~24 shows a linear polarization pattern  which
extends in both galaxies for several kiloparsecs. 
We use the 2D velocity field of ionized gas  in RR~24 members to interpret 
the linear polarization structure. RR~24a has quite regular gas kinematics. The gas velocity gradient in the North and South-West direction of the RR~24b nucleus matches the ``S-shape'' polarization pattern. 
We suggest that the  large-scale magnetic field of the RR~24 pair 
still  plays a role in shaping such a polarization patterns. 
We suggest that late-type galaxy magnetic fields may 
be distorted by interaction. The observed pattern may depend 
on the interaction stage.}

\end{itemize}

The case of RR~24 suggests a  coherence among the gas 
kinematics, the structure of the polarization and possibly of the
large-scale magnetic field. We believe that this pilot study underlines the 
need for a combined study of gas velocity field and polarization 
structure to infer  the evolution of the large-scale magnetic field stucture.  

We are performing a campaign of  polarization measures of nearby
galaxy pairs with the 1.82m Asiago Telescope using the AFOSC camera
equipped with a Double Wallaston Prism \citep{Desidera2002}. 
We intend also to obtain the 2D galaxy velocity fields using the 
GH$\alpha$FaS Fabry-Perot \citep{Hernandez08} at the 4.2m 
William Herschel Telescope at IAC in Canary Islands, Spain.

\section*{Acknowledgments}

This research has made use of the NASA/IPAC Extragalactic Database (NED) which
is operated by the Jet Propulsion Laboratory, California Institute of
Technology, under contract with the National Aeronautics and Space
Administration. The Digitized Sky Survey (DSS) was produced at the Space
Telescope Science Institute under U.S. Government grant NAG W-2166.


\end{document}